\documentclass[10pt,aps,prb,raggedbottom,longbibliography,nobalancelastpage,reprint,citeautoscript,letterpaper,superscriptaddress,fleqn]{revtex4-2}
\usepackage{graphicx}
\usepackage{mathrsfs}
\usepackage{amsfonts}
\usepackage{times}
\usepackage{amsmath}
\usepackage{color}
\usepackage[colorlinks,linkcolor=blue,citecolor=blue]{hyperref}
\usepackage{mathtools}
\usepackage{orcidlink}
\usepackage{multirow}
\usepackage{bm}
\usepackage{siunitx}
\usepackage{braket}
\usepackage{ulem}

\begin{document}
\title{
  Finite-Temperature {\it ab initio} Structural Optimization of the Bilayer Nickelate Superconductor 
  \texorpdfstring{La$_3$Ni$_2$O$_7$}{La3Ni2O7}
}

\author{Ryoma Asai}
\email{asai-ryoma238@g.ecc.u-tokyo.ac.jp}
\affiliation{Department of Applied Physics, The University of Tokyo, Hongo, Bunkyo-ku, Tokyo 113-8656, Japan}

\author{Ryotaro Arita}
\affiliation{Department of Physics, The University of Tokyo, Hongo, Bunkyo-ku, Tokyo 113-0033, Japan}
\affiliation{RIKEN Center for Emergent Matter Science, Hirosawa, Wako, Saitama 351-0198, Japan}

\author{Takumi Chida}
\affiliation{Department of Physics, The University of Tokyo, Hongo, Bunkyo-ku, Tokyo 113-0033, Japan}

\author{Ryota Masuki}
\affiliation{Department of Physics, The University of Tokyo, Hongo, Bunkyo-ku, Tokyo 113-0033, Japan}

\author{Kazuhiko Kuroki}
\affiliation{Department of Physics, The University of Osaka, Machikaneyama, Toyonaka, Osaka 560-0043, Japan}

\author{Terumasa Tadano}
\email{Tadano.Terumasa@nims.go.jp}
\affiliation{Research Center for Magnetic and Spintronic Materials, National Institute for Materials Science, Sengen, Tsukuba, Ibaraki 305-0047, Japan}
\begin{abstract}
We develop a first-principles framework for finite-temperature structural optimization that incorporates vibrational contributions to the free energy through anharmonic phonon theory. We extend and further improve the efficiency of the recent approach, enabling its application to systems in which the size of the primitive cell changes across structural phase transitions. Applying this framework to La$_3$Ni$_2$O$_7$, we establish its pressure-temperature phase diagram and find that the slope of the phase boundary between the high-symmetry and low-symmetry phases is negative, with a magnitude of approximately \SI{-60}{\kelvin\per\giga\pascal}. The present results provide a theoretical foundation for discussing how changes in crystal symmetry influence the emergence of superconductivity.
\end{abstract}

\maketitle
\textit{Introduction.}
The discovery of superconductivity with a transition temperature approaching \SI{80}{\kelvin} in the bilayer nickelate La$_3$Ni$_2$O$_7$ under pressure~\cite{sun2023signatures,zhang2024high,Hou_2023,PhysRevX.14.011040,PhysRevLett.131.126001,PhysRevB.111.174506,lechermann2023electronic,Shen_2023,christiansson2023correlated,shilenko2023correlated,liu2024electronic,wu2024superexchange,cao2024flat,chen2025charge,lu2024interlayer,oh2023type,liao2023electron,qu2024bilayer,PhysRevB.110.024514,jiang2024high,PhysRevB.95.214509,sakakibara2024possible,Wang2024,Shi2025,Yang2024,PhysRevLett.133.096002,doi:10.7566/JPSJ.94.054704,Xia2025,dygc-94fq,96K_Nature} has had a tremendous impact on the field of high-temperature superconductivity. While the pairing mechanism is not yet fully understood, it is almost certain that the contribution of the Ni $3d_{3z^2-r^2}$ orbital to the low-energy electronic states plays a crucial role. Because the $3d_{3z^2-r^2}$ orbital extends toward the apical oxygen sites, the Ni–O–Ni bond angle along the interlayer direction is expected to have a decisive influence on the emergence of superconductivity. These considerations suggest that the interplay between orbital character and crystallographic symmetry is one of the key issues to unravel the physics of superconductivity in La$_3$Ni$_2$O$_7$~\cite{PhysRevB.108.L140505,liu2023s,zhang2024structural,geisler2024structural}.

In order to elucidate the relation between crystal structure and superconductivity in La$_3$Ni$_2$O$_7$, a detailed understanding of its pressure–temperature ($p$–$T$) phase diagram is indispensable. Experimental studies have revealed that at ambient pressure the system crystallizes in the $Amam$ = $Cmcm$ (No. 63) structure below room temperature~\cite{ZHANG1994402,ling2000neutron,VORONIN2001202,doi:10.1143/JPSJ.64.1644}. Upon applying external pressure, structural transitions occur toward higher-symmetry phases such as $Fmmm$ (No. 69) or $I4/mmm$ (No. 139). Importantly, superconductivity has been observed only when the crystal structure possesses a symmetry higher than that of the $Amam$ phase.

This observation that higher symmetry favors superconductivity is further supported by recent thin-film experiments~\cite{Zhou2025_LaPr3Ni2O7,Ko2025_La3Ni2O7,Lorenz_2016,Middey2016_Nickelates,Belviso2019_AtomicDesign,Geisler2021_ThermoelectricOxideFilms,10.1093/nsr/nwaf205}. In epitaxial films, the in-plane lattice constants are reduced due to substrate constraints, which effectively increase the Ni-O-Ni bond angle toward 180$^{\circ}$. Under these conditions, superconductivity has been reported even at ambient pressure. These findings indicate that controlling the bond geometry through either pressure or epitaxial strain can decisively influence the stability of the superconducting phase.

Although these experimental results strongly indicate the critical role of structural symmetry, its $p$–$T$ phase diagram of La$_3$Ni$_2$O$_7$ is far from being fully established. In particular, the details of the phase diagram, such as which structural phases are realized at given $p$ and $T$ and where the phase boundaries lie, remain unsettled~\cite{wang2024structure,Wang2025_La3Ni2O7}. 
Thus, a first-principles determination of the phase diagram is essential to establish phase stability and its relation to superconductivity in La$_3$Ni$_2$O$_7$.

To address this problem, one must determine the $p$–$T$ phase diagram through structural optimization at finite temperatures. Such an optimization requires minimizing the free energy, rather than the total energy, incorporating the contributions of phonons. Recently, Masuki {\it et al.} developed a scheme that enables such finite-temperature structural optimization~\cite{PhysRevB.106.224104,PhysRevB.110.094102}. Their method has been successfully applied to the three-step structural phase transitions of BaTiO$_3$ as well as to the transitions of polar metals Li$B$O$_3$ ($B$=Ta, W, Re, Os), yielding results in good agreement with experiments.

However, their approach still faced challenges when applied to transitions accompanied by a modification of the unit cell, and convergence was often slow near phase boundaries. In this work, we overcome these difficulties by extending and accelerating the optimization scheme, and apply it to La$_3$Ni$_2$O$_7$. As a result, we establish its $p$–$T$ phase diagram, where the phase boundary between the high-symmetry $I4/mmm$ phase and the low-symmetry $Amam$ phase is found to form an almost straight line connecting ($p$,$T$) $\sim$ (\SI{0}{\giga\pascal}, \SI{630}{\kelvin}) and (\SI{10}{\giga\pascal}, \SI{0}{\kelvin}). These results provide a solid theoretical basis for understanding the emergence of superconductivity in La$_3$Ni$_2$O$_7$, where lattice symmetry plays a decisive role.

\textit{Method.}
We determine the pressure–temperature phase diagram of La$_3$Ni$_2$O$_7$ based on the self-consistent phonon (SCP) theory~\cite{Hooton01011958,gillis1968properties,tadano2022first,SOUVATZIS2009888,PhysRevB.105.064112,PhysRevB.92.054301,Monacelli_2021,PhysRevB.106.224104,PhysRevB.107.134119,PhysRevB.110.094102}. The general formulation of SCP-based structural optimization has been described in Ref.~\cite{PhysRevB.106.224104}; here we summarize the essential steps relevant to our study.

The SCP theory is an efficient mean-field approach for treating lattice anharmonicity, i.e., phonon--phonon interactions, and has been widely used to calculate finite-temperature phonons. In this framework, the lattice vibration under the anharmonic interaction $\hat{U}=\sum_{n=0}^{\infty}\hat{U}_n$, where $\hat{U}_n$ denotes the $n$-th order anharmonic term, is approximated by an effective harmonic (one-body) phonon Hamiltonian. As a trial Hamiltonian, we employ $\hat{\mathcal{H}}_{0}
    = \sum_{k\lambda} \hbar \Omega_{k\lambda} 
      \left( \hat{a}^{\dagger}_{k\lambda}\hat{a}_{k\lambda} + \frac{1}{2} \right)$,
where $\Omega_{k\lambda}$ are variational phonon frequencies.
The corresponding variational free energy is a functional of the trial Hamiltonian,
$F_{\mathrm{SCP}}[\hat{\mathcal{H}}_{0}]= \mathcal{F}_{0} + \bigl\langle \hat{T} + \hat{U} - \hat{\mathcal{H}}_{0} \bigr\rangle_{0},$
and satisfies the variational inequality $F \le F_{\mathrm{SCP}}[\hat{\mathcal{H}}_{0}]$.
Here, $\hat{T}$ is the kinetic-energy operator, 
$\mathcal{F}_0 = -\beta^{-1}\ln Z_0$, 
$\langle X\rangle_0 = Z_0^{-1}\mathrm{Tr}\!\left(X e^{-\beta\hat{\mathcal{H}}_{0}}\right)$ denotes the thermal average, 
and $Z_0=\mathrm{Tr}\!\left(e^{-\beta\hat{\mathcal{H}}_{0}}\right)$ is the corresponding partition function.

The optimal effective Hamiltonian is obtained by minimizing this functional, $\hat{\mathcal{H}}_{0}^{\mathrm{opt}}
    = \arg\min_{\hat{\mathcal{H}}_{0}} F_{\mathrm{SCP}}[\hat{\mathcal{H}}_{0}],
$
which we perform efficiently in reciprocal space by iteratively updating $\hat{\mathcal{H}}_{0}$ 
using interatomic force constants (IFCs) computed for a supercell~\cite{PhysRevB.92.054301}. 
Once $\hat{\mathcal{H}}_{0}^{\mathrm{opt}}$ is determined at each temperature, 
the corresponding value $F_{\mathrm{SCP}}[\hat{\mathcal{H}}_{0}^{\mathrm{opt}}]$ 
provides an accurate approximation of the exact vibrational free energy $F$, 
which is then used for structural and thermodynamic optimization.

To update atomic structures at finite temperatures within SCP theory, we use the analytical expression for the gradient of $F_{\mathrm{SCP}}[\hat{\mathcal{H}}_{0}^{\mathrm{opt}}]$ with respect to the atomic displacements $q_{\lambda}$, as derived by Masuki \textit{et al.}~\cite{PhysRevB.106.224104}. After each structure update, the IFCs must be updated to evaluate $F_{\mathrm{SCP}}[\hat{\mathcal{H}}_{0}^{\mathrm{opt}}]$ and its gradient for the new configuration. These IFC updates are efficiently performed using the ``IFC renormalization'' technique, which has been demonstrated to achieve reliable accuracy for various materials, including BaTiO$_3$~\cite{PhysRevB.106.224104}, ZnO, and GaN~\cite{PhysRevB.107.134119}.

\begin{figure}
    \centering
    \includegraphics[width=8.6cm,clip]{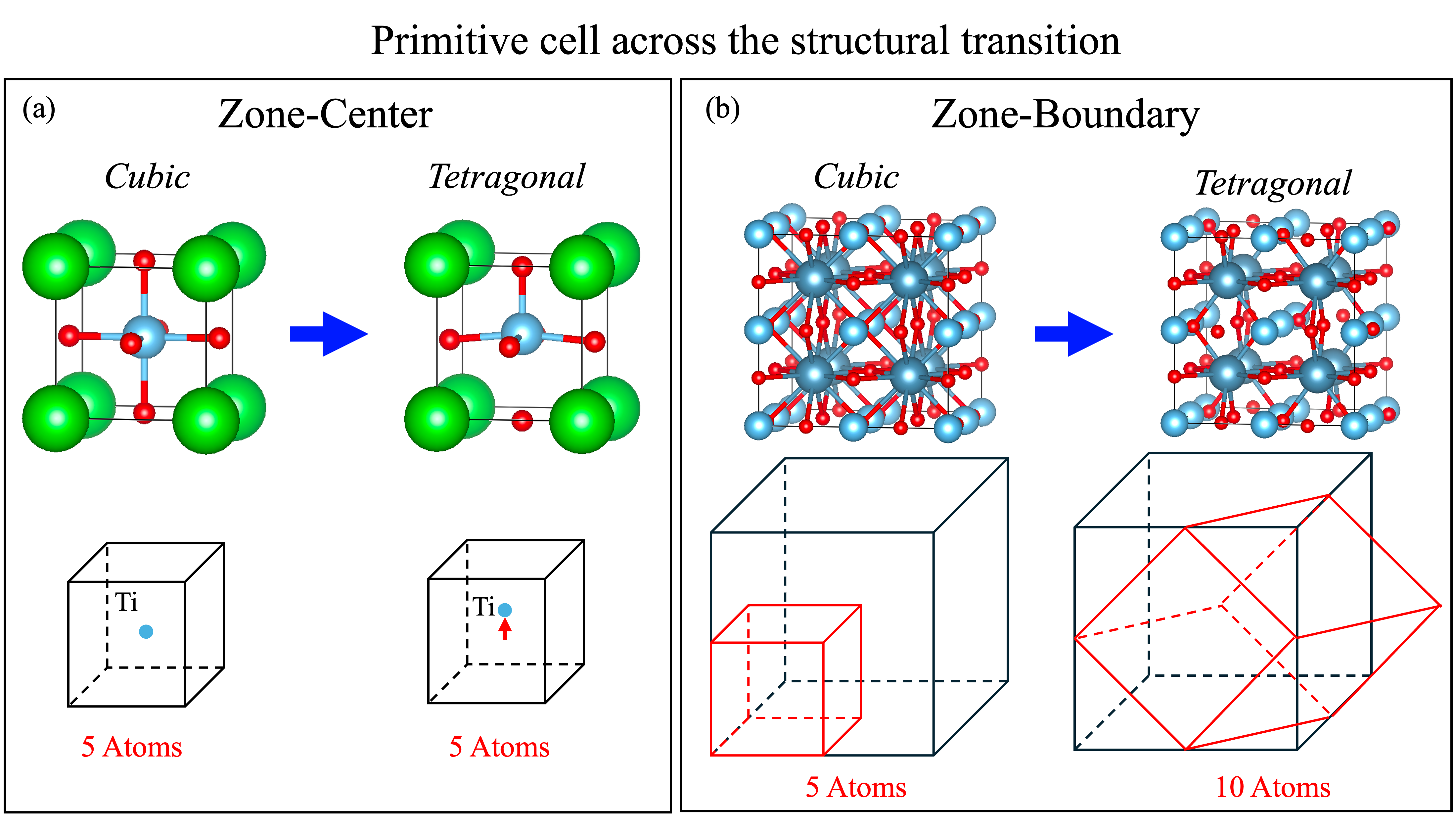}
    \caption{
    Schematic figures representing two types of structural phase transitions:
    (a) BaTiO$_3$, whose primitive cell size remains unchanged before and after the structural phase transition, and (b) CaTiO$_3$, whose primitive cell size changes. The crystal structures were visualized using VESTA\cite{Momma:db5098}.}
    \label{fig:Method}
\end{figure}

The original SCP-based structure optimization algorithm successfully reproduced phase transitions driven by zone-center ($\bm{q}= \bm{0}$) soft phonons (Fig.~\ref{fig:Method}(a)). However, it could not capture transitions induced by soft phonons at nonzero wavevectors ($\bm{q}\neq \bm{0}$) that involve changes in unit-cell size (Fig.~\ref{fig:Method}(b)). This limitation arose from the original implementation, which always performed the reciprocal-space updates of $\hat{\mathcal{H}}_{0}$ and $F_{\mathrm{SCP}}[\hat{\mathcal{H}}_{0}]$ using a primitive lattice. In this study, we have overcome this limitation by enabling the code to use a supercell throughout an SCP calculation. By choosing a supercell so that it becomes commensurate with the relevant soft mode, the $\bm{q}\neq \bm{0}$ soft mode in the primitive Brillouin zone is folded to the $\Gamma$-point of the supercell, allowing us to reuse the existing formalism for computing the gradient of $F_{\mathrm{SCP}}[\hat{\mathcal{H}}_{0}^{\mathrm{opt}}]$.

Although this extension appears straightforward, the increase in the cell size introduces numerical challenges, particularly slower convergence of the structure parameters when using the previously adopted Newton's method. To address this, we have improved the optimization algorithm by incorporating the elements of the GDIIS (Geometry optimization using Direct Inversion in the Iterative Subspace) scheme~\cite{B108658H,10.1063/1.477393,10.1063/1.480484}. This method couples the rapid RMM-DIIS  (Residual Minimization Method with Direct Inversion in the Iterative Subspace) electronic minimization~\cite{https://doi.org/10.1002/jcc.540030413} with a BFGS (Broyden–Fletcher–Goldfarb–Shanno) step~\cite{10.1093/imamat/6.1.76}, achieving robust convergence on smooth potential-energy surfaces~\cite{chida2024algorithm}.

Electronic-structure calculations were performed using the \textit{Vienna Ab initio Simulation Package} (VASP)~\cite{PhysRevB.54.11169}.  
We employed the PBE exchange--correlation functional~\cite{PhysRevLett.77.3865} and PAW pseudopotentials~\cite{PhysRevB.50.17953,PhysRevB.59.1758}.  
The SCP loop convergence criterion was set to $10^{-8}$~eV, and the accurate precision mode was used to suppress numerical errors and ensure reliable forces.  
For structural optimization of the 24-atom cell, a $12 \times 12 \times 3$ Monkhorst--Pack $k$ mesh and a 600~eV plane-wave cutoff were used. 
For the 96-atom supercell, a $7 \times 7 \times 3$ $k$ mesh with the same cutoff was employed.  
Although GGA+$U$ calculations~\cite{geisler2024structural,zhang2024structural} are typically employed in nickelate systems to achieve consistency with the Fermi level,  
because GGA+$U$ systematically underestimates lattice constants and predicts unrealistic transition temperatures, the main results presented here are based on GGA ($U=0$).  

The second-order IFCs were obtained from phonon calculations using a $2 \times 2 \times 1$ supercell containing 96 atoms.
To obtain the third- and fourth-order IFCs, 80 random configurations were generated by {\it ab initio} molecular dynamics (AIMD) at \SI{50}{\kelvin} with fixed lattice constants, combined with additional random displacements up to 0.04 \text{Bohr}.
The AIMD simulations were carried out with a $2 \times 2 \times 2$ $k$ mesh and a 400~eV cutoff. 
The cutoff radii of 10 \text{Bohr} and 7 \text{Bohr} were used for the third- and fourth-order IFCs, respectively.  
The IFCs were fitted using the LASSO method, yielding a fitting error of 3.55\%.
The SCP calculations were performed using the $2 \times 2 \times 1$ supercell with a $1 \times 1 \times 1$ $k$-point mesh. This cell size is large enough to accommodate the atomic distortions associated with the $I4/mmm$ to $Amam$ phase transition.
The SCP structural optimization at each temperature was iterated until the maximum atomic displacement became smaller than $10^{-4}$~\text{Bohr}.

\begin{figure}
    \centering
    \includegraphics[width=8.5cm, clip]{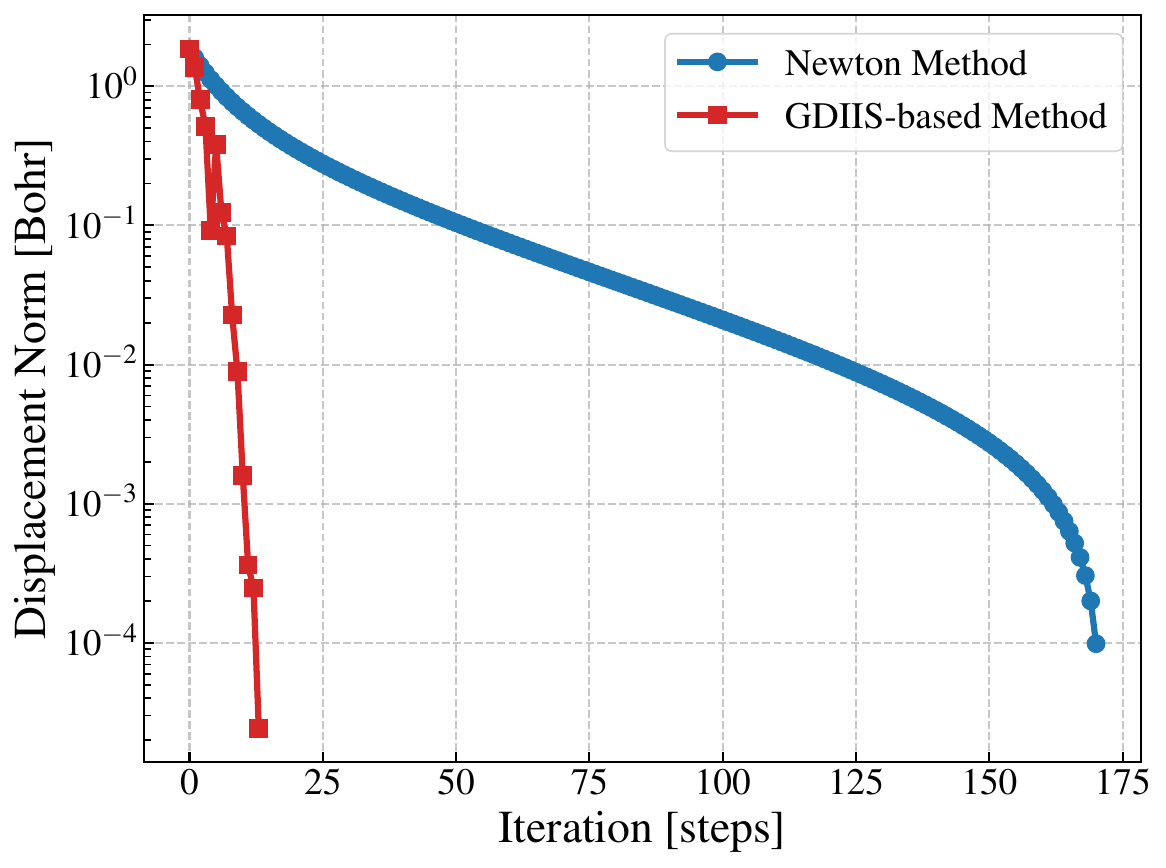}
    \caption{Iterative convergence behavior of two optimization schemes at \SI{0}{\giga\pascal} and \SI{380}{\kelvin}, showing the norm of atomic displacements versus iteration number.
}
    \label{fig:scheme}
\end{figure}

To benchmark the newly implemented GDIIS-based algorithm, we performed structural optimizations at \SI{0}{\giga\pascal} using two optimization schemes. We adopted the high-symmetry $I4/mmm$ structure as the initial configuration, slightly perturbed atoms in the direction of the $Amam$ phase, and started the structure optimization at \SI{380}{K}. Since the $Amam$ phase was more stable at this temperature, the magnitude of the displacement developed and eventually converged, as shown in Fig.~\ref{fig:scheme}. 
Here, we evaluated the atomic displacement norm relative to the final optimized structure, where the convergence criterion was defined such that the maximum atomic displacement is less than $10^{-4}$ Bohr.
It is clear that the GDIIS-based method accelerates the convergence by one order of magnitude, demonstrating its efficiency in structural optimizations at finite temperatures. We therefore employed the GDIIS-based scheme for all subsequent structural optimizations.

\textit{Results.} 
Before performing finite-temperature optimizations, we first investigated the structural phase transition at zero temperature ($T$=0). 
Although several experimental studies have suggested that the high-symmetry phase belongs to the $Fmmm$ space group~\cite{sun2023signatures,wang2024structure,Wang2025_La3Ni2O7}, 
theoretical works have reported that $Fmmm$ does not appear and instead transitions into $I4/mmm$~\cite{geisler2024structural,Ouyang2024}. 
Therefore, in this study, we focused on the phase boundary between $Amam$ and $I4/mmm$ and examined the critical pressure $p_\mathrm{s}$.

\begin{figure}
    \centering
    \includegraphics[width=8.5cm,clip]{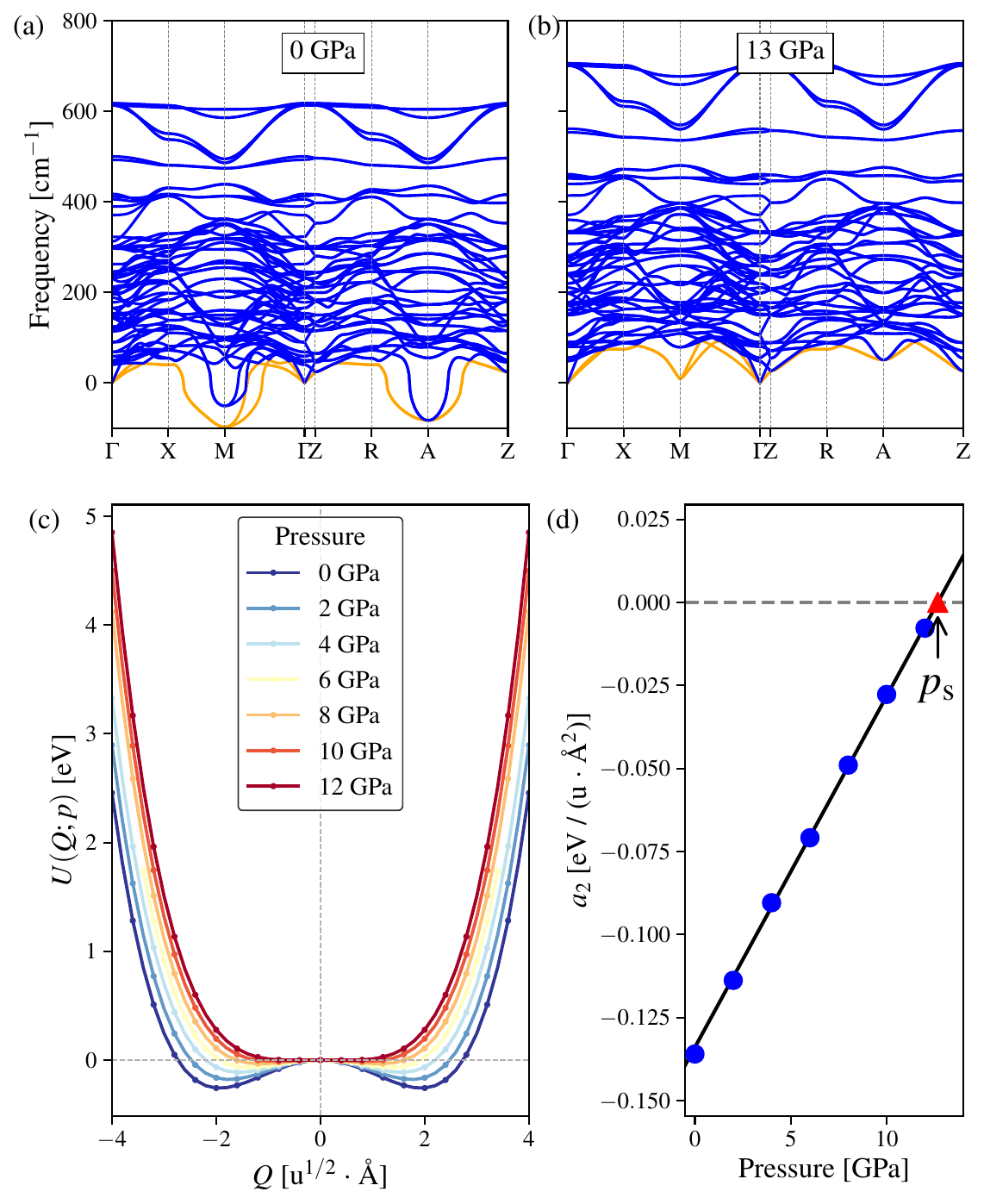}
    \caption{
    Phonon dispersion of the conventional cell of La$_3$Ni$_2$O$_7$ under different pressure conditions: (a) \SI{0}{\giga\pascal} and (b) \SI{13}{\giga\pascal}. The two lowest-energy modes at the M point are highlighted in orange.
    (c) Potential energy surface (PES) along the M-point modes at each pressure. 
    As the pressure increases, the double-well potential becomes shallower and approaches a single-well shape. 
    (d) Coefficients $a_2$ obtained by fitting the PES at each pressure with the quartic function $F(Q) = a_2 Q^2 + b_4 Q^4$.  We performed linear fit of $a_2(p)$, and the intercept on the pressure axis (red triangle) was taken as the estimated critical pressure $p_s$.
    }
    \label{fig:Phonon}
\end{figure}

Figures \ref{fig:Phonon}(a) and \ref{fig:Phonon}(b) display the phonon dispersions at $p_{0}=$ \SI{0}{\giga\pascal} and \SI{13}{\giga\pascal}, respectively. Here, $p_0$ represents a pressure of the static DFT calculation. 
Imaginary phonon modes are present around the M and A points at \SI{0}{\giga\pascal}, but these modes become stable at \SI{13}{\giga\pascal}. 
Figure \ref{fig:Phonon}(c) shows the potential energy surface (PES) along the normal coordinates $Q$ corresponding to the unstable M-point mode. 
At ambient pressure, the PES exhibits a double-well structure, which changes to single-well under compression. 
Based on these PES profiles, we estimated the transition pressure. 
Following the Landau theory~\cite{Landau:1937obd}, the PES is expressed as a function of the mode amplitude $Q$, and the potential energy is expanded up to the fourth order as $F(Q) = a_2Q^2 + b_4Q^4$. 
For each pressure, we fitted the PES to this quartic function and plotted the coefficients $a_2$ in Fig.\ref{fig:Phonon}~(d). 
Assuming, as in the Landau theory of phase transitions, that $a_2(p) = A(p - p_\mathrm{s})$, we performed a linear fit of $a_2(p)$ and obtained $p_\mathrm{s}$ from the intercept on the pressure axis. 
From this analysis, the critical pressure was estimated to be $p_\mathrm{s}=$ \SI{12.6}{\giga\pascal}.

\begin{figure}
    \centering
    \includegraphics[width=8.5cm,clip]{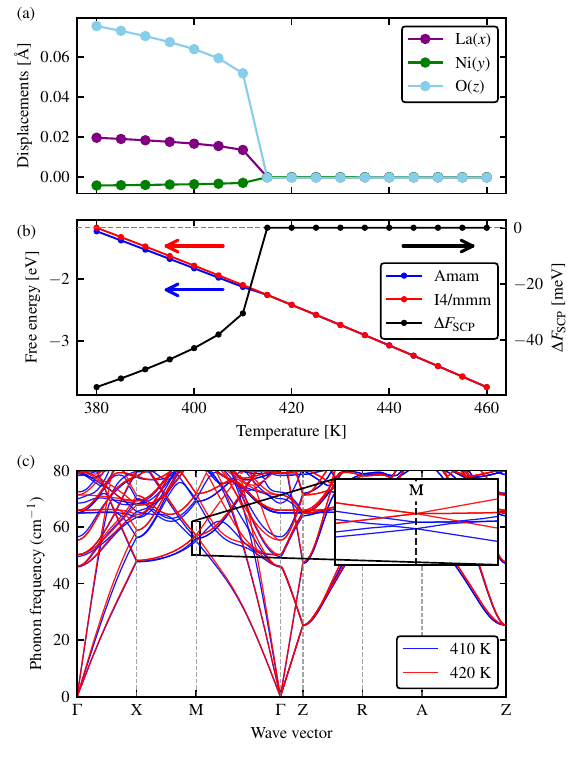}
    \caption{
    (a) Temperature dependence of atomic displacements obtained from finite-temperature structural optimization starting from the relaxed $I4/mmm$ structure. The onset of finite displacements indicates the symmetry change to $Amam$.
(b) Temperature dependence of the SCP free energy. The free-energy difference $\Delta F_{\mathrm{SCP}}$ between the $Amam$ (blue) and $I4/mmm$ (red) phases indicates that the former becomes thermodynamically stable when $\Delta F_{\mathrm{SCP}}<0$.
(c) Phonon band dispersions at $T=410$ and \SI{420}{\kelvin}, showing the splitting near the M point associated with the transition.
}
    \label{fig:displacements}
\end{figure}

Then we performed finite-temperature structural optimization based on the SCP theory. 
To reduce the computational cost associated with large supercells, the relaxation was restricted to the internal atomic coordinates while the lattice constants were fixed at the DFT-optimized values. 
We carried out SCP-based structural optimization at $p_0=$ \SI{0}{\giga\pascal} over a range of temperatures. 
Figure~\ref{fig:displacements}(a) shows the temperature dependence of the atomic displacements from the high-symmetry positions. 
At high temperatures, the system remains close to the reference $I4/mmm$ structure, while below a structural transition temperature, significant displacements appear. 
The pattern of atomic displacements—namely, in-plane shifts of La and Ni atoms and out-of-plane displacements of O atoms—corresponds to the octahedral rotation characteristic of the $Amam$ phase. 
A symmetry analysis using spglib~\cite{Togo31122024} revealed a clear symmetry reduction from $I4/mmm$ to $Amam$ at the onset temperature. 
Furthermore, as shown in Fig.~\ref{fig:displacements}(b), the total SCP free energy of the $Amam$ structure becomes lower than that of the $I4/mmm$ phase at the same temperature, confirming the thermodynamic stability of the distorted phase. 
From the onset of atomic displacements, the emergence of symmetry breaking, and the crossing of free energies, we determine that the finite-temperature phase transition occurs at \SI{411}{\kelvin} under $p_0=$ \SI{0}{\giga\pascal}.

Figure~\ref{fig:displacements}(c) shows the phonon band structures calculated from the SCP frequencies at \SI{0}{\giga\pascal}. 
The spectra at \SI{410}{\kelvin} and \SI{420}{\kelvin}, just below and above the transition temperature, are overlaid for comparison. 
Focusing on the M-point modes, one can clearly see that the degeneracy present at high temperature is lifted in the low-temperature spectrum, indicating the symmetry breaking associated with the structural phase transition.

\begin{figure}
    \centering
    \includegraphics[width=8.5cm, clip]{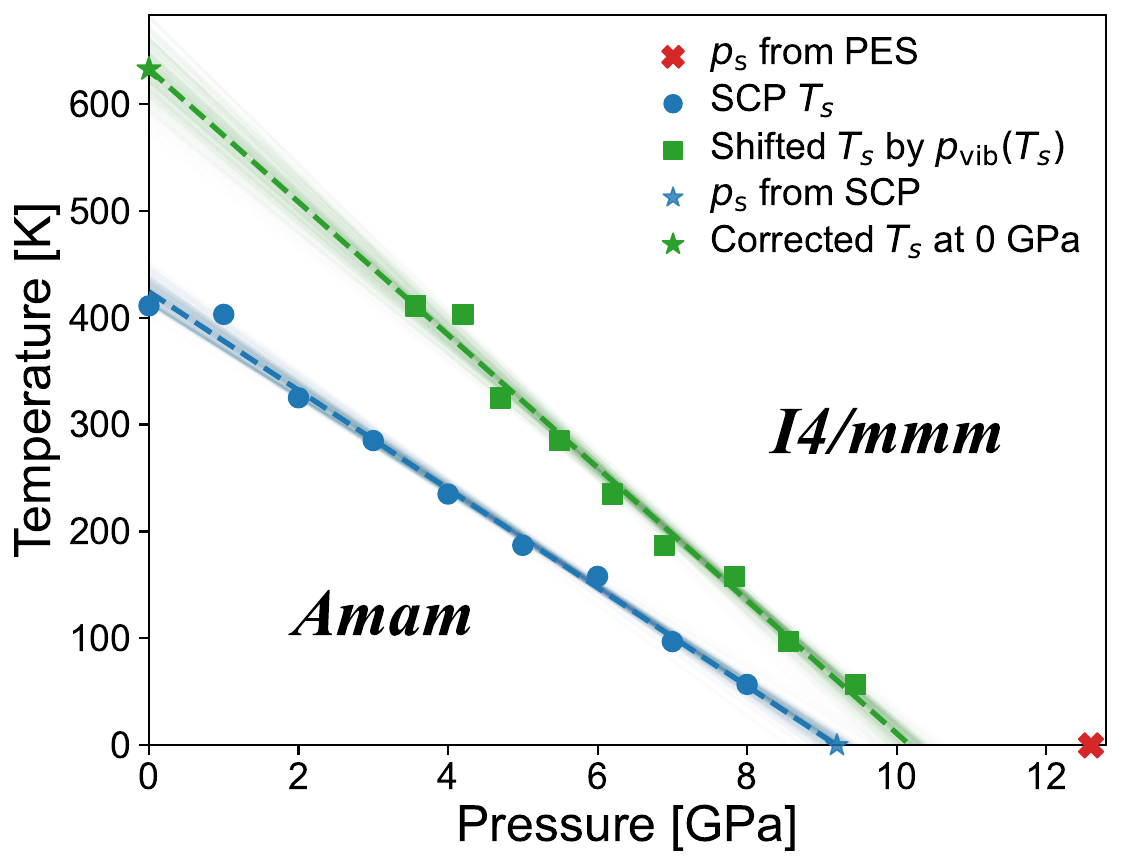}
    \caption{
   Phase diagram of La$3$Ni$2$O$7$ obtained from self-consistent phonon (SCP) calculations.
Blue points denote the transition temperatures $T\mathrm{s}$ calculated under fixed pressures from 0 to \SI{8}{\giga\pascal}.
The blue dashed line represents a linear fit obtained using bootstrap resampling, and the shaded region indicates the corresponding confidence interval. The intercept of this fit with the $T=$ \SI{0}{\kelvin} axis provides the transition pressure $p\mathrm{s}$ estimated from SCP calculations.
The red triangle indicates the transition pressure $p\mathrm{s}$ determined from potential energy surface.
Green points show the $T_\mathrm{s}$ values corrected by the vibrational pressure $P_\mathrm{vib}(T_\mathrm{s})$, which accounts for the effect of finite-temperature lattice vibrations.
The green dashed line corresponds to a bootstrap-based linear fit to the $p_\mathrm{vib}$-corrected data, and the shaded band gives its confidence range. The intercept of this line with the $p=$ \SI{0}{\giga\pascal} axis yields the corrected transition temperature $T_\mathrm{s}$ at \SI{0}{\giga\pascal}.}.
    \label{fig:phase_diagram}
\end{figure}

For each pressure condition, we repeated the structural optimization based on the SCP theory, 
while keeping the lattice constants fixed to the DFT-optimized values at the corresponding pressure. 
For each pressure, we extracted the $T_s$ at which the $Amam$-type distortion becomes stable. 
The resulting temperature–pressure phase diagram is shown in Fig.~\ref{fig:phase_diagram} (blue points). 
These $T_s$ values obtained from the SCP calculations at various $p_{0}$ can be well approximated by a linear function, 
from which the phase boundary between the $Amam$ and $I4/mmm$ phases can be estimated. More specifically, the phase boundary is found to form an almost straight line connecting ($p_0$,$T$) $\sim$ (\SI{0}{\giga\pascal}, \SI{400}{\kelvin}) and (\SI{9}{\giga\pascal}, \SI{0}{\kelvin}). The estimated critical pressure of $\sim$ \SI{9}{\giga\pascal} at \SI{0}{K} is smaller than $p_\mathrm{s}=$ \SI{12.6}{\giga\pascal} obtained from the PES at \SI{0}{K} (red triangle). This difference can be partially attributed to zero-point vibration, which tends to stabilize the high-symmetry $I4/mmm$ phase.

Next, we correct the phase boundary by considering the effect of phonon excitations on pressure. The vibrational contribution to pressure, which is not included in the static hydrostatic pressure $p_0$, was evaluated as follows.
\begin{equation}
p_{\mathrm{vib}}(T)
= -\frac{\partial F_{\mathrm{vib}}}{\partial \Omega}
= \frac{1}{\Omega}\sum_{\bm{q}j}\hbar\omega_{\bm{q}j}
\left(n_{\bm{q}j}+\frac{1}{2}\right)\gamma_{\bm{q}j}.
\end{equation}
Here, $\Omega$ is the unit-cell volume, $\omega_{\bm{q}j}$ is the SCP frequency at temperature $T$, and $\gamma_{\bm{q}j}$ is the corresponding Grüneisen parameter. 
By shifting the pressure by $p_{\mathrm{vib}}(T_s)$, we obtained the corrected phase boundary (green points). 
When this correction is applied, the phase boundary shifts toward higher pressures, and the resulting $T_s(p)$ curve can still be well approximated by a linear function. 
The extrapolated y-intercept is $\sim$630 \si{K}, which is close to the experimental result \SI{720}{\kelvin}\cite{D0TA06731H}, indicating that the present approach yields quantitatively reasonable results.

\textit{Conclusion.}
We have predicted the $p$–$T$ phase diagram of the superconducting compound La$_3$Ni$_2$O$_7$ by performing finite-temperature structural optimization based on the SCP theory.
Our calculations reveal that the phase boundary between the low- and high-symmetry phases does not run vertically in the pressure axis, but instead exhibits a significant temperature dependence even at ambient pressure. This indicates that a structural phase transition can occur purely as a function of temperature, without applying external pressure.

More broadly, we expect that the ability to perform finite-temperature structural optimization will enable reliable predictions of $p$–$T$ phase diagrams in a wide variety of materials. The method holds promise as a general theoretical framework to guide the discovery of new functional materials exhibiting structural or electronic phase transitions under external conditions.

\textit{Acknowledgments.}
This work is supported by Grant-in-Aid for Scientific Research from JSPS, KAKENHI Grant  No. 25H01246, No. 25H01252, No. 24H00190, JST K-Program JPMJKP25Z7, JST-PRESTO JPMJPR23J6, RIKEN TRIP initiative (RIKEN Quantum, Advanced General Intelligence for Science Program, Many-body Electron Systems).

\bibliography{main.bib}

@article{sun2023signatures,
  title   = {Signatures of superconductivity near 80 {K} in a nickelate under high pressure},
  author  = {Sun, Hualei and Huo, Mengwu and Hu, Xunwu and Li, Jingyuan and Liu, Zengjia and Han, Yifeng and Tang, Lingyun and Mao, Zhongquan and Yang, Pengtao and Wang, Bosen},
  journal = {Nature},
  volume  = {621},
  number  = {7979},
  pages   = {493--498},
  year    = {2023},
  doi     = {10.1038/s41586-023-06408-7}
}

@article{geisler2024structural,
  title   = {Structural transitions, octahedral rotations, and electronic properties of {$\text{A}_{3}\text{Ni}_{2}\text{O}_{7}$} rare-earth nickelates under high pressure},
  author  = {Geisler, Benjamin and Hamlin, James J and Stewart, Gregory R and Hennig, Richard G and Hirschfeld, P.~J.},
  journal = {npj Quantum Mater.},
  volume  = {9},
  number  = {1},
  pages   = {38},
  year    = {2024},
  doi     = {10.1038/s41535-024-00648-0}
}

@article{wang2024structure,
  title   = {Structure responsible for the superconducting state in {$\text{La}_{3}\text{Ni}_{2}\text{O}_{7}$} at high-pressure and low-temperature conditions},
  author  = {Wang, Luhong and Li, Yan and Xie, Sheng-Yi and Liu, Fuyang and Sun, Hualei and Huang, Chaoxin and Gao, Yang and Nakagawa, Takeshi and Fu, Boyang and Dong, Bo},
  journal = {J. Am. Chem. Soc.},
  volume  = {146},
  number  = {11},
  pages   = {7506--7514},
  year    = {2024},
  doi     = {10.1021/jacs.3c13094}
}

@article{Hou_2023,
  title   = {Emergence of High-Temperature Superconducting Phase in Pressurized {$\text{La}_{3}\text{Ni}_{2}\text{O}_{7}$} Crystals},
  author  = {Hou, Jun and Yang, Peng-Tao and Liu, Zi-Yi and Li, Jing-Yuan and Shan, Peng-Fei and Ma, Liang and Wang, Gang and Wang, Ning-Ning and Guo, Hai-Zhong and Sun, Jian-Ping and Uwatoko, Yoshiya and Wang, Meng and Zhang, Guang-Ming and Wang, Bo-Sen and Cheng, Jin-Guang},
  journal = {Chin. Phys. Lett.},
  volume  = {40},
  number  = {11},
  pages   = {117302},
  year    = {2023},
  doi     = {10.1088/0256-307X/40/11/117302}
}

@article{zhang2024high,
  title   = {High-temperature superconductivity with zero resistance and strange-metal behaviour in {$\text{La}_{3}\text{Ni}_{2}\text{O}_{7-\delta}$}},
  author  = {Zhang, Yanan and Su, Dajun and Huang, Yanen and Shan, Zhaoyang and Sun, Hualei and Huo, Mengwu and Ye, Kaixin and Zhang, Jiawen and Yang, Zihan and Xu, Yongkang},
  journal = {Nat. Phys.},
  volume  = {20},
  number  = {8},
  pages   = {1269--1273},
  year    = {2024},
  doi     = {10.1038/s41567-024-02515-y}
}

@article{PhysRevLett.131.126001,
  title   = {Bilayer Two-Orbital Model of {$\text{La}_{3}\text{Ni}_{2}\text{O}_{7}$} under Pressure},
  author  = {Luo, Zhihui and Hu, Xunwu and Wang, Meng and W\'u, W\'ei and Yao, Dao-Xin},
  journal = {Phys. Rev. Lett.},
  volume  = {131},
  number  = {12},
  pages   = {126001},
  year    = {2023},
  doi     = {10.1103/PhysRevLett.131.126001}
}

@article{PhysRevB.111.174506,
  title   = {Effective model and pairing tendency in the bilayer Ni-based superconductor {$\text{La}_{3}\text{Ni}_{2}\text{O}_{7}$}},
  author  = {Gu, Yuhao and Le, Congcong and Yang, Zhesen and Wu, Xianxin and Hu, Jiangping},
  journal = {Phys. Rev. B},
  volume  = {111},
  number  = {17},
  pages   = {174506},
  year    = {2025},
  doi     = {10.1103/PhysRevB.111.174506}
}

@article{PhysRevB.108.L140505,
  title   = {Possible ${s}_{\ifmmode\pm\else\textpm\fi{}}$-wave superconductivity in {$\text{La}_{3}\text{Ni}_{2}\text{O}_{7}$}},
  author  = {Yang, Qing-Geng and Wang, Da and Wang, Qiang-Hua},
  journal = {Phys. Rev. B},
  volume  = {108},
  number  = {14},
  pages   = {L140505},
  year    = {2023},
  doi     = {10.1103/PhysRevB.108.L140505}
}

@article{lechermann2023electronic,
  title   = {Electronic correlations and superconducting instability in {$\text{La}_{3}\text{Ni}_{2}\text{O}_{7}$} under high pressure},
  author  = {Lechermann, Frank and Gondolf, Jannik and B\"otzel, Steffen and Eremin, Ilya M.},
  journal = {Phys. Rev. B},
  volume  = {108},
  number  = {20},
  pages   = {L201121},
  year    = {2023},
  doi     = {10.1103/PhysRevB.108.L201121}
}

@article{sakakibara2024possible,
  title   = {Possible High ${T}_{c}$ Superconductivity in {$\text{La}_{3}\text{Ni}_{2}\text{O}_{7}$} under High Pressure through Manifestation of a Nearly Half-Filled Bilayer Hubbard Model},
  author  = {Sakakibara, Hirofumi and Kitamine, Naoya and Ochi, Masayuki and Kuroki, Kazuhiko},
  journal = {Phys. Rev. Lett.},
  volume  = {132},
  number  = {10},
  pages   = {106002},
  year    = {2024},
  doi     = {10.1103/PhysRevLett.132.106002}
}

@article{Shen_2023,
  title   = {Effective Bi-Layer Model Hamiltonian and Density-Matrix Renormalization Group Study for the High-Tc Superconductivity in {$\text{La}_{3}\text{Ni}_{2}\text{O}_{7}$} under High Pressure},
  author  = {Shen, Yang and Qin, Mingpu and Zhang, Guang-Ming},
  journal = {Chin. Phys. Lett.},
  volume  = {40},
  number  = {12},
  pages   = {127401},
  year    = {2023},
  doi     = {10.1088/0256-307X/40/12/127401}
}

@article{christiansson2023correlated,
  title   = {Correlated Electronic Structure of {$\text{La}_{3}\text{Ni}_{2}\text{O}_{7}$} under Pressure},
  author  = {Christiansson, Viktor and Petocchi, Francesco and Werner, Philipp},
  journal = {Phys. Rev. Lett.},
  volume  = {131},
  number  = {20},
  pages   = {206501},
  year    = {2023},
  doi     = {10.1103/PhysRevLett.131.206501}
}

@article{shilenko2023correlated,
  title   = {Correlated electronic structure, orbital-selective behavior, and magnetic correlations in double-layer {$\text{La}_{3}\text{Ni}_{2}\text{O}_{7}$} under pressure},
  author  = {Shilenko, D.~A. and Leonov, I.~V.},
  journal = {Phys. Rev. B},
  volume  = {108},
  number  = {12},
  pages   = {125105},
  year    = {2023},
  doi     = {10.1103/PhysRevB.108.125105}
}

@article{liu2024electronic,
  title   = {Electronic correlations and partial gap in the bilayer nickelate {$\text{La}_{3}\text{Ni}_{2}\text{O}_{7}$}},
  author  = {Liu, Zhe and Huo, Mengwu and Li, Jie and Li, Qing and Liu, Yuecong and Dai, Yaomin and Zhou, Xiaoxiang and Hao, Jiahao and Lu, Yi and Wang, Meng},
  journal = {Nat. Commun.},
  volume  = {15},
  number  = {1},
  pages   = {7570},
  year    = {2024},
  doi     = {10.1038/s41467-024-52001-5}
}

@article{wu2024superexchange,
  title   = {Superexchange and charge transfer in the nickelate superconductor {$\text{La}_{3}\text{Ni}_{2}\text{O}_{7}$} under pressure},
  author  = {W{\'u}, W{\'e}i and Luo, Zhihui and Yao, Dao-Xin and Wang, Meng},
  journal = {Sci. China Phys. Mech. Astron.},
  volume  = {67},
  number  = {11},
  pages   = {117402},
  year    = {2024},
  doi     = {10.1007/s11433-023-2300-4}
}

@article{cao2024flat,
  title   = {Flat bands promoted by Hund's rule coupling in the candidate double-layer high-temperature superconductor {$\text{La}_{3}\text{Ni}_{2}\text{O}_{7}$} under high pressure},
  author  = {Cao, Yingying and Yang, Yi-feng},
  journal = {Phys. Rev. B},
  volume  = {109},
  number  = {8},
  pages   = {L081105},
  year    = {2024},
  doi     = {10.1103/PhysRevB.109.L081105}
}

@article{chen2025charge,
  title   = {Charge and spin instabilities in superconducting {$\text{La}_{3}\text{Ni}_{2}\text{O}_{7}$}},
  author  = {Chen, Xuejiao and Jiang, Peiheng and Li, Jie and Zhong, Zhicheng and Lu, Yi},
  journal = {Phys. Rev. B},
  volume  = {111},
  number  = {1},
  pages   = {014515},
  year    = {2025},
  doi     = {10.1103/PhysRevB.111.014515}
}

@article{liu2023s,
  title   = {${s}^{\ifmmode\pm\else\textpm\fi{}}$-Wave Pairing and the Destructive Role of Apical-Oxygen Deficiencies in {$\text{La}_{3}\text{Ni}_{2}\text{O}_{7}$} under Pressure},
  author  = {Liu, Yu-Bo and Mei, Jia-Wei and Ye, Fei and Chen, Wei-Qiang and Yang, Fan},
  journal = {Phys. Rev. Lett.},
  volume  = {131},
  number  = {23},
  pages   = {236002},
  year    = {2023},
  doi     = {10.1103/PhysRevLett.131.236002}
}

@article{lu2024interlayer,
  title   = {Interlayer-Coupling-Driven High-Temperature Superconductivity in {$\text{La}_{3}\text{Ni}_{2}\text{O}_{7}$} under Pressure},
  author  = {Lu, Chen and Pan, Zhiming and Yang, Fan and Wu, Congjun},
  journal = {Phys. Rev. Lett.},
  volume  = {132},
  number  = {14},
  pages   = {146002},
  year    = {2024},
  doi     = {10.1103/PhysRevLett.132.146002}
}

@article{zhang2024structural,
  title   = {Structural phase transition, s$_{\pm}$-wave pairing, and magnetic stripe order in bilayered superconductor {$\text{La}_{3}\text{Ni}_{2}\text{O}_{7}$} under pressure},
  author  = {Zhang, Yang and Lin, Ling-Fang and Moreo, Adriana and Maier, Thomas A and Dagotto, Elbio},
  journal = {Nat. Commun.},
  volume  = {15},
  number  = {1},
  pages   = {2470},
  year    = {2024},
  doi     = {10.1038/s41467-024-46622-z}
}

@article{oh2023type,
  title   = {Type-II $t\ensuremath{-}J$ model and shared superexchange coupling from Hund's rule in superconducting {$\text{La}_{3}\text{Ni}_{2}\text{O}_{7}$}},
  author  = {Oh, Hanbit and Zhang, Ya-Hui},
  journal = {Phys. Rev. B},
  volume  = {108},
  number  = {17},
  pages   = {174511},
  year    = {2023},
  doi     = {10.1103/PhysRevB.108.174511}
}

@article{liao2023electron,
  title   = {Electron correlations and superconductivity in {$\text{La}_{3}\text{Ni}_{2}\text{O}_{7}$} under pressure tuning},
  author  = {Liao, Zhiguang and Chen, Lei and Duan, Guijing and Wang, Yiming and Liu, Changle and Yu, Rong and Si, Qimiao},
  journal = {Phys. Rev. B},
  volume  = {108},
  number  = {21},
  pages   = {214522},
  year    = {2023},
  doi     = {10.1103/PhysRevB.108.214522}
}

@article{qu2024bilayer,
  title   = {Bilayer ${t\text{\ensuremath{-}}J\text{\ensuremath{-}}J}_{\ensuremath{\perp}}$ Model and Magnetically Mediated Pairing in the Pressurized Nickelate {$\text{La}_{3}\text{Ni}_{2}\text{O}_{7}$}},
  author  = {Qu, Xing-Zhou and Qu, Dai-Wei and Chen, Jialin and Wu, Congjun and Yang, Fan and Li, Wei and Su, Gang},
  journal = {Phys. Rev. Lett.},
  volume  = {132},
  number  = {3},
  pages   = {036502},
  year    = {2024},
  doi     = {10.1103/PhysRevLett.132.036502}
}

@article{Wang2025_La3Ni2O7,
  title   = {Temperature-Dependent Structural Evolution of Ruddlesden--Popper Bilayer Nickelate {$\text{La}_{3}\text{Ni}_{2}\text{O}_{7}$}},
  author  = {Wang, Haozhe and Zhou, Haidong and Xie, Weiwei},
  journal = {Inorg. Chem.},
  volume  = {64},
  number  = {2},
  pages   = {828--834},
  year    = {2025},
  doi     = {10.1021/acs.inorgchem.4c03042}
}

@article{ZHANG1994402,
  title   = {Synthesis, Structure, and Properties of the Layered Perovskite {$\text{La}_{3}\text{Ni}_{2}\text{O}_{7-\delta}$}},
  author  = {Zhang, Z. and Greenblatt, M. and Goodenough, J.~B.},
  journal = {J. Solid State Chem.},
  volume  = {108},
  number  = {2},
  pages   = {402--409},
  year    = {1994},
  doi     = {10.1006/jssc.1994.1059}
}

@article{ling2000neutron,
  title   = {Neutron Diffraction Study of {$\text{La}_{3}\text{Ni}_{2}\text{O}_{7}$}: Structural Relationships Among $n=1, 2,$ and $3$ Phases $\text{La}_{n+1}\text{Ni}_{n}\text{O}_{3n+1}$},
  author  = {Ling, Christopher and Argyriou, Dimitri and Wu, Guoqing and Neumeier, J.},
  journal = {J. Solid State Chem.},
  volume  = {152},
  pages   = {517--525},
  year    = {2000},
  doi     = {10.1006/jssc.2000.8721}
}

@article{VORONIN2001202,
  title   = {Neutron diffraction, synchrotron radiation and EXAFS spectroscopy study of crystal structure peculiarities of the lanthanum nickelates $\text{La}_{n+1}\text{Ni}_{n}\text{O}_{y}$ ($n=1,2,3$)},
  author  = {Voronin, V.~I. and Berger, I.~F. and Cherepanov, V.~A. and Gavrilova, L.~Ya. and Petrov, A.~N. and Ancharov, A.~I. and Tolochko, B.~P. and Nikitenko, S.~G.},
  journal = {Nucl. Instrum. Methods Phys. Res., Sect. A},
  volume  = {470},
  number  = {1},
  pages   = {202--209},
  year    = {2001},
  doi     = {10.1016/S0168-9002(01)01036-1}
}

@article{doi:10.1143/JPSJ.64.1644,
  title   = {Transport, Magnetic and Thermal Properties of {$\text{La}_{3}\text{Ni}_{2}\text{O}_{7-\delta}$}},
  author  = {Taniguchi, Satoshi and Nishikawa, Takashi and Yasui, Yukio and Kobayashi, Yoshiaki and Takeda, Jun and Shamoto, Shin-ichi and Sato, Masatoshi},
  journal = {J. Phys. Soc. Jpn.},
  volume  = {64},
  number  = {5},
  pages   = {1644--1650},
  year    = {1995},
  doi     = {10.1143/JPSJ.64.1644}
}

@article{Lorenz_2016,
  title   = {The 2016 oxide electronic materials and oxide interfaces roadmap},
  author  = {Lorenz, M. and Ramachandra Rao, M.~S. and Venkatesan, T. and Fortunato, E. and Barquinha, P. and Branquinho, R. and Salgueiro, D. and Martins, R. and Carlos, E. and Liu, A.},
  journal = {J. Phys. D: Appl. Phys.},
  volume  = {49},
  number  = {43},
  pages   = {433001},
  year    = {2016},
  doi     = {10.1088/0022-3727/49/43/433001}
}

@article{Middey2016_Nickelates,
  title   = {Physics of Ultrathin Films and Heterostructures of Rare-Earth Nickelates},
  author  = {Middey, S. and Chakhalian, J. and Mahadevan, P. and Freeland, J.~W. and Millis, A.~J. and Sarma, D.~D.},
  journal = {Annu. Rev. Mater. Res.},
  volume  = {46},
  pages   = {305--334},
  year    = {2016},
  doi     = {10.1146/annurev-matsci-070115-032057}
}

@article{Belviso2019_AtomicDesign,
  title   = {Viewpoint: Atomic-Scale Design Protocols toward Energy, Electronic, Catalysis, and Sensing Applications},
  author  = {Belviso, Florian and Claerbout, Victor E.~P. and Comas-Vives, Aleix and Dalal, Naresh S. and Fan, Feng-Ren and Filippetti, Alessio and Fiorentini, Vincenzo and Foppa, Lucas and Franchini, Cesare and Geisler, Benjamin},
  journal = {Inorg. Chem.},
  volume  = {58},
  number  = {22},
  pages   = {14939--14980},
  year    = {2019},
  doi     = {10.1021/acs.inorgchem.9b01785}
}

@article{Geisler2021_ThermoelectricOxideFilms,
  title   = {Tuning the Thermoelectric Properties of Transition Metal Oxide Thin Films and Superlattices on the Quantum Scale},
  author  = {Geisler, Benjamin and Yordanov, Petar and Gruner, Markus Ernst and Keimer, Bernhard and Pentcheva, Rossitza},
  journal = {Phys. Status Solidi B},
  volume  = {259},
  number  = {12},
  pages   = {2100270},
  year    = {2021},
  doi     = {10.1002/pssb.202100270}
}

@article{Zhou2025_LaPr3Ni2O7,
  title   = {Ambient-pressure superconductivity onset above 40 K in {$(\text{La},\text{Pr})_{3}\text{Ni}_{2}\text{O}_{7}$} films},
  author  = {Zhou, Guangdi and Lv, Wei and Wang, Heng and Nie, Zihao and Chen, Yaqi and Li, Yueying and Huang, Haoliang and Chen, Wei-Qiang and Sun, Yu-Jie and Xue, Qi-Kun},
  journal = {Nature},
  volume  = {640},
  number  = {8059},
  pages   = {641--646},
  year    = {2025},
  doi     = {10.1038/s41586-025-08755-z}
}

@article{Ko2025_La3Ni2O7,
  title   = {Signatures of ambient pressure superconductivity in thin film {$\text{La}_{3}\text{Ni}_{2}\text{O}_{7}$}},
  author  = {Ko, Eun Kyo and Yu, Yijun and Liu, Yidi and Bhatt, Lopa and Li, Jiarui and Thampy, Vivek and Kuo, Cheng-Tai and Wang, Bai Yang and Lee, Yonghun and Lee, Kyuho},
  journal = {Nature},
  volume  = {638},
  number  = {8052},
  pages   = {935--940},
  year    = {2025},
  doi     = {10.1038/s41586-024-08525-3}
}

@article{Hooton01011958,
  title   = {The use of a model in anharmonic lattice dynamics},
  author  = {Hooton, D.~J.},
  journal = {Philos. Mag.},
  volume  = {3},
  number  = {25},
  pages   = {49--54},
  year    = {1958},
  doi     = {10.1080/14786435808243224}
}

@article{gillis1968properties,
  title   = {Properties of Crystalline Argon and Neon in the Self-Consistent Phonon Approximation},
  author  = {Gillis, N.~S. and Werthamer, N.~R. and Koehler, T.~R.},
  journal = {Phys. Rev.},
  volume  = {165},
  number  = {3},
  pages   = {951--959},
  year    = {1968},
  doi     = {10.1103/PhysRev.165.951}
}

@article{tadano2022first,
  title   = {First-Principles Phonon Quasiparticle Theory Applied to a Strongly Anharmonic Halide Perovskite},
  author  = {Tadano, Terumasa and Saidi, Wissam A.},
  journal = {Phys. Rev. Lett.},
  volume  = {129},
  number  = {18},
  pages   = {185901},
  year    = {2022},
  doi     = {10.1103/PhysRevLett.129.185901}
}

@article{SOUVATZIS2009888,
  title   = {The self-consistent ab initio lattice dynamical method},
  author  = {Souvatzis, P. and Eriksson, O. and Katsnelson, M.~I. and Rudin, S.~P.},
  journal = {Comput. Mater. Sci.},
  volume  = {44},
  number  = {3},
  pages   = {888--894},
  year    = {2009},
  doi     = {10.1016/j.commatsci.2008.06.016}
}

@article{PhysRevB.105.064112,
  title   = {Anharmonic Gr\"uneisen theory based on self-consistent phonon theory: Impact of phonon-phonon interactions neglected in the quasiharmonic theory},
  author  = {Masuki, Ryota and Nomoto, Takuya and Arita, Ryotaro and Tadano, Terumasa},
  journal = {Phys. Rev. B},
  volume  = {105},
  number  = {6},
  pages   = {064112},
  year    = {2022},
  doi     = {10.1103/PhysRevB.105.064112}
}

@article{PhysRevB.92.054301,
  title   = {Self-consistent phonon calculations of lattice dynamical properties in cubic {$\text{SrTiO}_{3}$} with first-principles anharmonic force constants},
  author  = {Tadano, Terumasa and Tsuneyuki, Shinji},
  journal = {Phys. Rev. B},
  volume  = {92},
  number  = {5},
  pages   = {054301},
  year    = {2015},
  doi     = {10.1103/PhysRevB.92.054301}
}

@article{PhysRevB.106.224104,
  title   = {Ab initio structural optimization at finite temperatures based on anharmonic phonon theory: Application to the structural phase transitions of {${\mathrm{BaTiO}}_{3}$}},
  author  = {Masuki, Ryota and Nomoto, Takuya and Arita, Ryotaro and Tadano, Terumasa},
  journal = {Phys. Rev. B},
  volume  = {106},
  number  = {22},
  pages   = {224104},
  year    = {2022},
  doi     = {10.1103/PhysRevB.106.224104}
}

@article{chida2024algorithm,
  title   = {Algorithm Improvement for Finite-Temperature Structural Optimization Using Anharmonic Phonon Theory},
  author  = {Chida, Takumi and Masuki, Ryota},
  journal = {Unpublished report},
  year    = {2024},
  url     = {https://www.merit.t.u-tokyo.ac.jp/merit/en/training/pdf/report/jihatsu_chida_masuki_eng.pdf}
}

@article{D0TA06731H,
  title   = {Structure, electrical conductivity and oxygen transport properties of Ruddlesden--Popper phases $\text{Ln}_{n+1}\text{Ni}_{n}\text{O}_{3n+1}$ {(Ln = La, Pr and Nd; n = 1, 2 and 3)}},
  author  = {Song, Jia and Ning, De and Boukamp, Bernard and Bassat, Jean-Marc and Bouwmeester, Henny J.~M.},
  journal = {J. Mater. Chem. A},
  volume  = {8},
  number  = {42},
  pages   = {22206--22221},
  year    = {2020},
  doi     = {10.1039/D0TA06731H}
}

@article{Togo31122024,
  title   = {Spglib: a software library for crystal symmetry search},
  author  = {Togo, Atsushi and Shinohara, Kohei and Tanaka, Isao},
  journal = {Sci. Technol. Adv. Mater.: Methods},
  volume  = {4},
  number  = {1},
  pages   = {2384822},
  year    = {2024},
  doi     = {10.1080/27660400.2024.2384822}
}

@article{PhysRevLett.77.3865,
  title   = {Generalized Gradient Approximation Made Simple},
  author  = {Perdew, John P. and Burke, Kieron and Ernzerhof, Matthias},
  journal = {Phys. Rev. Lett.},
  volume  = {77},
  number  = {18},
  pages   = {3865--3868},
  year    = {1996},
  doi     = {10.1103/PhysRevLett.77.3865}
}

@article{Monacelli_2021,
  title   = {The stochastic self-consistent harmonic approximation: calculating vibrational properties of materials with full quantum and anharmonic effects},
  author  = {Monacelli, Lorenzo and Bianco, Raffaello and Cherubini, Marco and Calandra, Matteo and Errea, Ion and Mauri, Francesco},
  journal = {J. Phys.: Condens. Matter},
  volume  = {33},
  number  = {36},
  pages   = {363001},
  year    = {2021},
  doi     = {10.1088/1361-648X/ac066b}
}

@article{PhysRevB.107.134119,
  title   = {Full optimization of quasiharmonic free energy with an anharmonic lattice model: Application to thermal expansion and pyroelectricity of wurtzite {$\text{GaN}$} and {$\text{ZnO}$}},
  author  = {Masuki, Ryota and Nomoto, Takuya and Arita, Ryotaro and Tadano, Terumasa},
  journal = {Phys. Rev. B},
  volume  = {107},
  number  = {13},
  pages   = {134119},
  year    = {2023},
  doi     = {10.1103/PhysRevB.107.134119}
}

@article{PhysRevB.54.11169,
  title   = {Efficient iterative schemes for ab initio total-energy calculations using a plane-wave basis set},
  author  = {Kresse, G. and Furthm\"uller, J.},
  journal = {Phys. Rev. B},
  volume  = {54},
  number  = {16},
  pages   = {11169--11186},
  year    = {1996},
  doi     = {10.1103/PhysRevB.54.11169}
}

@article{PhysRevB.50.17953,
  title   = {Projector augmented-wave method},
  author  = {Bl\"ochl, P.~E.},
  journal = {Phys. Rev. B},
  volume  = {50},
  number  = {24},
  pages   = {17953--17979},
  year    = {1994},
  doi     = {10.1103/PhysRevB.50.17953}
}

@article{PhysRevB.59.1758,
  title   = {From ultrasoft pseudopotentials to the projector augmented-wave method},
  author  = {Kresse, G. and Joubert, D.},
  journal = {Phys. Rev. B},
  volume  = {59},
  number  = {3},
  pages   = {1758--1775},
  year    = {1999},
  doi     = {10.1103/PhysRevB.59.1758}
}

@article{PhysRevX.14.011040,
  title   = {Pressure-Induced Superconductivity In Polycrystalline {$\text{La}_{3}\text{Ni}_{2}\text{O}_{7-\delta}$}},
  author  = {Wang, G. and Wang, N.~N. and Shen, X.~L. and Hou, J. and Ma, L. and Shi, L.~F. and Ren, Z.~A. and Gu, Y.~D. and Ma, H.~M. and Yang, P.~T.},
  journal = {Phys. Rev. X},
  volume  = {14},
  number  = {1},
  pages   = {011040},
  year    = {2024},
  doi     = {10.1103/PhysRevX.14.011040}
}

@article{https://doi.org/10.1002/jcc.540030413,
  title   = {Improved SCF convergence acceleration},
  author  = {Pulay, P.},
  journal = {J. Comput. Chem.},
  volume  = {3},
  number  = {4},
  pages   = {556--560},
  year    = {1982},
  doi     = {10.1002/jcc.540030413}
}

@article{10.1093/imamat/6.1.76,
  title   = {The Convergence of a Class of Double-rank Minimization Algorithms 1. General Considerations},
  author  = {Broyden, C.~G.},
  journal = {IMA J. Appl. Math.},
  volume  = {6},
  number  = {1},
  pages   = {76--90},
  year    = {1970},
  doi     = {10.1093/imamat/6.1.76}
}

@article{B108658H,
  title   = {Methods for optimizing large molecules. Part III. An improved algorithm for geometry optimization using direct inversion in the iterative subspace {(GDIIS)}},
  author  = {Farkas, {\"O}d{\"o}n and Schlegel, H. Bernhard},
  journal = {Phys. Chem. Chem. Phys.},
  volume  = {4},
  number  = {1},
  pages   = {11--15},
  year    = {2002},
  doi     = {10.1039/B108658H}
}

@article{10.1063/1.477393,
  title   = {Methods for geometry optimization of large molecules. I. An {$O(\text{N}^{2})$} algorithm for solving systems of linear equations for the transformation of coordinates and forces},
  author  = {Farkas, {\"O}d{\"o}n and Schlegel, H. Bernhard},
  journal = {J. Chem. Phys.},
  volume  = {109},
  number  = {17},
  pages   = {7100--7104},
  year    = {1998},
  doi     = {10.1063/1.477393}
}

@article{10.1063/1.480484,
  title   = {Methods for optimizing large molecules. II. Quadratic search},
  author  = {Farkas, {\"O}d{\"o}n and Schlegel, H. Bernhard},
  journal = {J. Chem. Phys.},
  volume  = {111},
  number  = {24},
  pages   = {10806--10814},
  year    = {1999},
  doi     = {10.1063/1.480484}
}

@article{PhysRevB.110.094102,
  title   = {Continuous crossover between insulating ferroelectrics and polar metals: Ab initio calculation of structural phase transitions of {Li$B$O$_3$ ($B$=Ta, W, Re, Os)}},
  author  = {Masuki, Ryota and Nomoto, Takuya and Arita, Ryotaro and Tadano, Terumasa},
  journal = {Phys. Rev. B},
  volume  = {110},
  number  = {9},
  pages   = {094102},
  year    = {2024},
  doi     = {10.1103/PhysRevB.110.094102}
}

@article{Momma:db5098,
  title   = {{\it VESTA3} for three-dimensional visualization of crystal, volumetric and morphology data},
  author  = {Momma, Koichi and Izumi, Fujio},
  journal = {J. Appl. Crystallogr.},
  volume  = {44},
  number  = {6},
  pages   = {1272--1276},
  year    = {2011},
  doi     = {10.1107/S0021889811038970}
}

@article{Landau:1937obd,
  title   = {On the theory of phase transitions},
  author  = {Landau, L.~D.},
  journal = {Zh. Eksp. Teor. Fiz.},
  volume  = {7},
  pages   = {19--32},
  year    = {1937}
}

@article{PhysRevB.110.024514,
  title   = {Superconductivity in nickelate and cuprate superconductors with strong bilayer coupling},
  author  = {Fan, Zhen and Zhang, Jian-Feng and Zhan, Bo and Lv, Dingshun and Jiang, Xing-Yu and Normand, Bruce and Xiang, Tao},
  journal = {Phys. Rev. B},
  volume  = {110},
  number  = {2},
  pages   = {024514},
  year    = {2024},
  doi     = {10.1103/PhysRevB.110.024514}
}

@article{jiang2024high,
  title   = {High-temperature superconductivity in {$\text{La}_{3}\text{Ni}_{2}\text{O}_{7}$}},
  author  = {Jiang, Kun and Wang, Ziqiang and Zhang, Fu-Chun},
  journal = {Chin. Phys. Lett.},
  volume  = {41},
  number  = {1},
  pages   = {017402},
  year    = {2024},
  doi     = {10.1088/0256-307X/41/1/017402}
}

@article{PhysRevB.95.214509,
  title   = {Finite-energy spin fluctuations as a pairing glue in systems with coexisting electron and hole bands},
  author  = {Nakata, Masahiro and Ogura, Daisuke and Usui, Hidetomo and Kuroki, Kazuhiko},
  journal = {Phys. Rev. B},
  volume  = {95},
  number  = {21},
  pages   = {214509},
  year    = {2017},
  doi     = {10.1103/PhysRevB.95.214509}
}

@article{Ouyang2024,
  title   = {Absence of electron-phonon coupling superconductivity in the bilayer phase of {$\text{La}_{3}\text{Ni}_{2}\text{O}_{7}$} under pressure},
  author  = {Ouyang, Zhenfeng and Gao, Miao and Lu, Zhong-Yi},
  journal = {npj Quantum Mater.},
  volume  = {9},
  number  = {1},
  pages   = {80},
  year    = {2024},
  doi     = {10.1038/s41535-024-00689-5}
}

@article{PhysRevLett.133.096002,
  title   = {Quenched Pair Breaking by Interlayer Correlations as a Key to Superconductivity in {$\text{La}_{3}\text{Ni}_{2}\text{O}_{7}$}},
  author  = {Ryee, Siheon and Witt, Niklas and Wehling, Tim O.},
  journal = {Phys. Rev. Lett.},
  volume  = {133},
  number  = {9},
  pages   = {096002},
  year    = {2024},
  doi     = {10.1103/PhysRevLett.133.096002}
}
\end{document}